\documentclass[10pt,aps,prb,twocolumn,preprintnumbers,amsmath,amssymb,floatfix,showpacs,citeautoscript,superscriptaddress%,linenumbers
]{revtex4-2}
\usepackage[utf8]{inputenc}
\usepackage[T1]{fontenc}
\usepackage[english]{babel}
\usepackage[pdftex]{graphicx}
\usepackage[normalem]{ulem}
\usepackage{multirow}
\usepackage{amsmath}
\usepackage{times}
\usepackage{color}
\usepackage[usenames,dvipsnames,svgnames,table]{xcolor}
\usepackage[colorlinks,plainpages=false,linkcolor=blue,urlcolor=blue,citecolor=blue,pdfpagemode=UseNone]{hyperref}
\usepackage{svg}

\renewcommand{\AA}{\text{\r{A}}}
\newcommand{\mrm}[1]{\mathrm{#1}}
\newcommand{\mbf}[1]{\mathbf{#1}}
\newcommand{\Tc}{T_{\mrm{c}}}

\newcommand{\wtwo}{\omega_{2}}

\newcommand{\atf}{\alpha^2F(\omega)}

\newcommand{\bk}{\mbf{k}}
\newcommand{\bq}{\mbf{q}}

\definecolor{d12orange}{rgb}{0.8500    0.3250    0.0980}
\definecolor{g8yellow}{rgb}{0.    0.6    0.298}
\definecolor{ppurple}{rgb}{0.4940    0.1840    0.5560}

\definecolor{mag}{RGB}{255,0,255}

\begin{document}

\title
{
\boldmath
Accelerating superconductor discovery through tempered deep learning of the electron-phonon spectral function
}

\author{Jason B. Gibson}
\email{jasongibson@ufl.edu} \thanks{These authors contributed equally to this work}
\affiliation{Department of Materials Science and Engineering, University of Florida, Gainesville, Florida 32611, USA}
\affiliation{Quantum Theory Project, University of Florida, Gainesville, Florida 32611, USA}

\author{Ajinkya C. Hire}
\email{ajinkya.hire@ufl.edu} \thanks{These authors contributed equally to this work}
\affiliation{Department of Materials Science and Engineering, University of Florida, Gainesville, Florida 32611, USA}
\affiliation{Quantum Theory Project, University of Florida, Gainesville, Florida 32611, USA}

\author{Philip M. Dee}
\affiliation{Department of Materials Science and Engineering, University of Florida, Gainesville, Florida 32611, USA}
\affiliation{Department of Physics, University of Florida, Gainesville, Florida 32611, USA}

\author{Oscar Barrera}
\affiliation{Department of Physics, University of Florida, Gainesville, Florida 32611, USA}

\author{Benjamin Geisler}
% \email{benjamin.geisler@ufl.edu}
\affiliation{Department of Materials Science and Engineering, University of Florida, Gainesville, Florida 32611, USA}
\affiliation{Department of Physics, University of Florida, Gainesville, Florida 32611, USA}

\author{Peter J. Hirschfeld}
% \email{pjh@phys.ufl.edu}
\affiliation{Department of Physics, University of Florida, Gainesville, Florida 32611, USA}
\author{Richard G. Hennig}
% \email{rhennig@ufl.edu}
\affiliation{Department of Materials Science and Engineering, University of Florida, Gainesville, Florida 32611, USA}
\affiliation{Quantum Theory Project, University of Florida, Gainesville, Florida 32611, USA}
\affiliation{Department of Physics, University of Florida, Gainesville, Florida 32611, USA}

\date{\today}

%==============================================================================%
%==============================================================================%
%==============================================================================%
\begin{abstract}
Integrating deep learning with the search for new electron-phonon superconductors represents a burgeoning field of research, where the primary challenge lies in the computational intensity of calculating the electron-phonon spectral function, $\atf$, the essential ingredient of Midgal-Eliashberg theory of superconductivity. To overcome this challenge, we adopt a two-step approach. First, we compute $\atf$ for 818 dynamically stable materials. We then train a deep-learning model to predict $\atf$, using an unconventional training strategy to temper the model's overfitting, enhancing predictions. Specifically,  we train a  Bootstrapped Ensemble of Tempered Equivariant graph neural NETworks (BETE-NET), obtaining an MAE of 0.21, 45 K, and 43 K for the Eliashberg moments derived from $\atf$: $\lambda$, $\omega_{\log}$, and $\wtwo$, respectively, yielding an MAE of 2.5 K for the critical temperature, $T_c$. Further, we incorporate domain knowledge of the site-projected phonon density of states to impose inductive bias into the model's node attributes and enhance predictions. This methodological innovation decreases the MAE to 0.18, 29 K, and 28 K, respectively, yielding an MAE of 2.1 K for $T_c$. We illustrate the practical application of our model in high-throughput screening for high-$\Tc$ materials. The model demonstrates an average precision nearly five times higher than random screening, highlighting the potential of ML in accelerating superconductor discovery. BETE-NET accelerates the search for high-$\Tc$ superconductors while setting a precedent for applying ML in materials discovery, particularly when data is limited.

\end{abstract}

\maketitle

\section{Introduction}

The world is currently undergoing an AI revolution that is having profound effects on science and society, brought about by the integration of predictive models. The foundational AI models driving this revolution consist of billions of model parameters trained on immense datasets.
In the realm of physical sciences, researchers frequently encounter a significant challenge employing such models: the datasets available are often inhomogeneous and limited in size. This scarcity of comprehensive datasets can substantially hinder the progress and accuracy of scientific discoveries~\cite{MGI}. Superconductivity, a field at the forefront of modern physics, exemplifies the small-dataset issue. Despite its transformative potential in areas such as energy transmission, magnetic levitation for transportation, and powerful superconducting magnets for medical imaging~\cite{Yao2021}, the development and understanding of new superconducting materials are often constrained by the paucity of large, comprehensive datasets. We tackle this quintessential problem of small datasets in the field of superconductivity by integrating physics directly into our models and leveraging the concept of the double descent phenomenon, which is integral to the success of foundational AI~\cite{double_descent_origin, double_descent}.

\begin{figure*}
    \centering
    \includegraphics[width=0.9\linewidth]{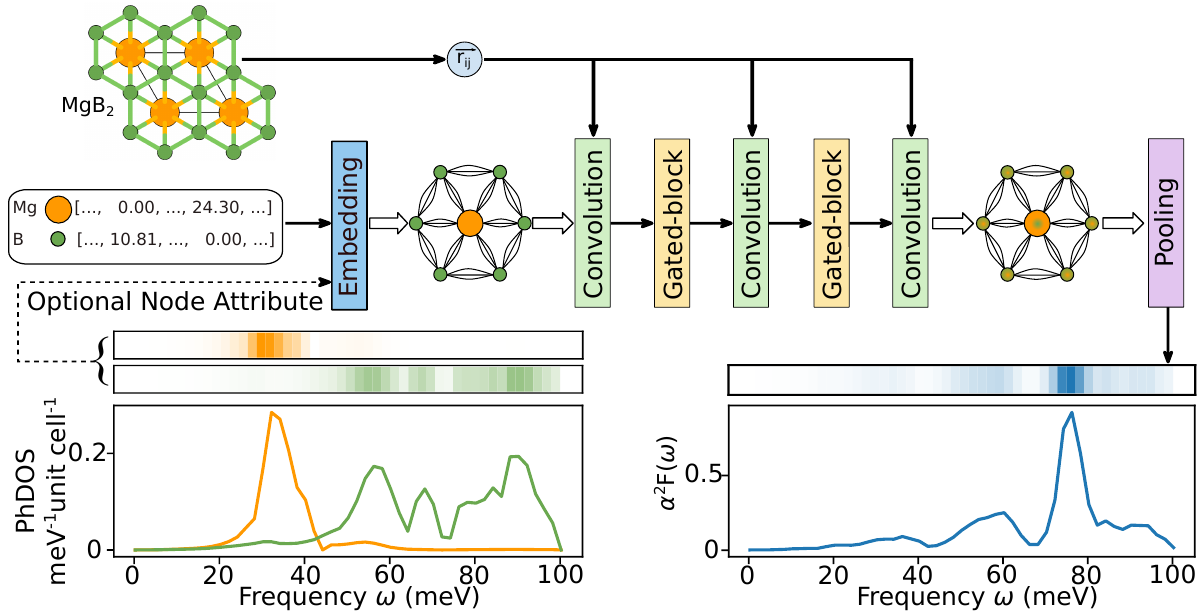}
    \caption{\textbf{BETE-NET architecture for predicting the Eliashberg function.} First, the model converts a crystal structure into a graph with nodes representing atoms. The nodes are embedded with a one-hot-encoded vector of the constituent atoms' atomic number multiplied by the atomic mass. The edges, connecting atoms within a 4~$\AA$ radius, embed the interatomic distance. A sequence of convolutions and gated-block operations are then applied to the graph. Finally, a pooling operation is applied to the node embedding, yielding the prediction of $\atf$. This is the base network for the crystal-structure-only (CSO) variant of BETE-NET. A bootstrapped ensemble of 100 of these models is then trained to produce the final model. The model's prediction is vastly improved by including information on the site-projected PhDOS. To enable this, we add a decision block to determine if the site-projected PhDOS is to be appended to the initial node embedding. There are two possible embeddings: the coarse PhDOS (CPD) model variant embeds coarse site-projected PhDOS (force-constant calculated on $2\times2\times2$ $\bq$-grid and Fourier interpolated to $20\times20\times20$) into the nodes. The fine PhDOS (FPD) model variant is the same as CPD but with a PhDOS embedding calculated on a fine $\bq$-grid and Fourier interpolated to $20\times20\times20$.}
    \label{network}
\end{figure*}

Machine learning (ML) has been massively successful in accelerating the discovery of materials by providing predictions that bypass the computationally expensive calculations required for determining thermodynamic stability~\cite{Wu2013, Jennings2019, SahinovicGeisler:21, SahinovicGeisler:22, Charraud2022, Li2023, Gibson2022} and characterizing materials~\cite{Xie_Tian2018, Ward2016}. ML can expedite the discovery of new potential electron-phonon superconductors by accelerating a strategy that has emerged since the theoretical predictions~\cite{Duan2014,Liu2017,Peng2017,Ashcroft2004} and subsequent discovery of high-temperature hydride superconductors~\cite{Drozdov_H3S_2015,somayazulu2019evidence,Drozdov2019}. Typically, one chooses a chemical system and identifies thermodynamically stable phases within it using crystal structure prediction algorithms coupled with density functional theory (DFT).  One then
calculates the Eliashberg spectral function (electron-phonon spectral function, $\atf$) for these stable or meta-stable materials, from which the superconducting critical temperature, $\Tc$, is calculated either by using the Allen-Dynes equation~\cite{Allen-Dynes1975}, Xie equation~\cite{Xie2022} or the more accurate Migdal-Eliashberg theory~\cite{Migdal1958, Eliashberg1960, Eliashberg1961}.

%\PH{How we will solve world hunger here.}
Development of models for rapidly and accurately estimating the superconducting properties of metals has been hindered as the methods used for evolving the aforementioned models~\cite{Wu2013, Jennings2019, Charraud2022, Li2023, Gibson2022, Xie_Tian2018, Ward2016} require tens of thousands of data points from materials informatics databases~\cite{Jain2013, Kirklin2015, Curtarolo2012, Choudhary2020}. Unlike these databases, correspondingly large datasets of $\atf$ are challenging to develop both because of the prohibitive cost and because of the lack of a standardized set of DFT parameters - for example, $\bk$-point and $\bq$-point density, smearing values - for accurately calculating $\atf$. In light of these obstacles, there is a need for ML techniques that can effectively work with small datasets.

Past works~\cite{Zeng2019, Li2020, Roter2020, Konno2021, Kim-Dordevic2022, Stanev2018} have addressed the hurdle of limited data for superconducting properties by utilizing the well-known "SuperCon" database~\cite{supercon2011} comprising experimental $\Tc$ values. However, this database is rife with repeated entries, questionable values, and unclear chemical formulas~\cite{Stanev2018, Hamidieh2018, Kim-Dordevic2022, Sommer2022_3DSC}. These inconsistencies and the paucity of supplementary material information have led some groups to build alternative databases~\cite{Hosono2015, Sommer2022_3DSC, Foppiano2023}. Recently, databases of material structures and calculated $\atf$ have emerged as well~\cite{Superhydra, Hoffmann2023, Choudhary2022, Cerqueira2023}. The Superhydra database~\cite{Superhydra} consists of only high-pressure hydrides, and the database by Hoffmann \emph{et al.}~\cite{Hoffmann2023} focuses on Heusler superconductors. Cerqueira \emph{et al.}~\cite{Cerqueira2023} recently performed 7000 electron-phonon calculations and trained a model on this data for simultaneously predicting electron-phonon coupling constant, $\lambda$, the logarithmic moment, $\omega_{\text{log}}$, of $\frac{2}{\lambda\omega}\atf$, and $\Tc$. While the work presented here was being conducted, this dataset was unavailable, but it has recently been made public~\cite{Cerqueira2023_database}. Choudary \emph{et al.}~\cite{Choudhary2022} developed a database of 626 dynamically stable materials with the associated $\atf$, which laid the foundation for a viable database for training a model to predict $\atf$. Choudary and co-workers then trained the ALIGNN~\cite{ALIGNN} to predict $\atf$. Unfortunately, 9\% of the materials exhibit negative values for $\alpha^2F(\omega)$ - an unphysical behavior - and nearly 8\% are duplicates, yielding only 521 entries for training, testing, and validation. 

The effectiveness of AI models for superconductors hinges on two key factors: the training dataset and the choice of machine learning technique. This work addresses these crucial elements by creating a comprehensive dataset of Eliashberg spectral functions and developing robust models using modern deep-learning techniques. Specifically, we propose an algorithm for standardizing the choice of $\bk$ and $\bq$-grids by generating grids based on user-provided $\bk$ and $\bq$-point densities, in contrast to using fixed grids for materials with different unit cell volumes. This method produces a comprehensive database comprising high-quality electron-phonon calculations for 818 dynamically stable materials. 

We then use this information to fit a linear regression model, which serves as the basis of comparison for our deep-learning models. This baseline model performs comparably to existing neural networks reported in the literature. Further, this simple model demonstrates the importance of using more than one error metric for comparing the performance of the models, as a single metric provides an incomplete evaluation.
 
To address the limited size of our database for deep learning, we design BETE-NET(Fig.~\ref{network}), using a modified version of the network proposed by Chen \emph{et al.}~\cite{Chen2021_ENN}.
For our target property, we prefer predicting $\alpha^2F(\omega)$ over directly predicting single-valued properties like Allen-Dynes $\Tc$, electron-phonon coupling constant ($\lambda$), and logarithmic moment ($\omega_{\text{log}}$), as the predicted $\alpha^2F(\omega)$ provides more insights into the workings and failures of the model than one gains from single-valued predictions. 
Further, directly predicting $\alpha^2F(\omega)$ also overcomes the hurdle of fixing a single value for the Coulomb repulsion ($\mu^*$) that works well for materials comprising transition and non-transition elements. Moreover, all of these properties, along with the superconducting gap and isotropic Eliashberg $\Tc$, can be derived from the more fundamental $\alpha^2F(\omega)$.
% Intriguingly, $\atf$ is of key relevance in a much broader range of applications; for instance, it plays a central role in understanding the relaxation kinetics of hot charge carriers in metals~\cite{Kratzer-RelaxationKinetics:22}.
Notably, the relevance of $\atf$ in applications extends beyond superconductivity and is instrumental in studying transport phenomena~\cite{Allen1971,Allen1978}. For instance, it is used to calculate the electrical resistivity of metals due to electron-phonon scattering and plays a central role in understanding the relaxation kinetics of hot charge carriers in metals~\cite{Kratzer-RelaxationKinetics:22}. 

We further delve into the nuances of training by allowing our models to train far beyond the onset of overfitting, revealing a double descent~\cite{double_descent} behavior that allows tempered overfitting~\cite{overfitting} of our model. While the details of the double descent phenomenon and tempered overfitting are beyond the scope of this paper, we provided a visual explanation in the results section and refer the readers to the work of Nakkiran \emph{et al.}\cite{overfitting,double_descent} for further details. 
%Rephrase next sentence
We incorporate physics-informed inductive bias by embedding the site-projected phonon density of states (PhDOS) in our model to improve our prediction results further. The performance of our models is summarized in Table~\ref{model_errors}. Notably, our model,  which takes the crystal structure as the input (referred to as CSO-crystal structure only) and was trained on just 651 examples, rivals the performance of models from existing literature trained on $\sim$7,000 datapoints~\cite{Cerqueira2023}.

Finally, we demonstrate the practical utility of our models by outlining a multi-step strategy to use both the no-cost CSO model and the moderately expensive coarse-PhDOS (CPD) model for high-throughput screening. By analyzing the precision-recall curves, we show the expected performance at each step, with our best model obtaining an average precision nearly five times that of random selection. Our strategy empowers researchers to prioritize promising materials for further investigation, accelerating the discovery of new electron-phonon superconductors.

\begin{table}
\caption{Comparison of the testing RMSE, MAE, and coefficient of determination, $\mrm{R}^2$, of the baseline, CSO, CPD, and FPD model for the electron-phonon coupling constant and the two moments of $\frac{2}{\lambda\omega}\atf$ -- logarithmic moment $\omega_{\text{log}}$ and second moment $\omega_2$. Incorporating domain knowledge about the PhDOS reduces the error.}
\begin{tabular}{c|c|ccc}
\hline
\multirow{2}{*}{Models} & \multirow{2}{*}{Metrics} & \multicolumn{3}{c}{Property}                                     \\
                        &                          & \multicolumn{1}{c|}{$\lambda$} & \multicolumn{1}{c|}{$\omega_{\text{log}}$ (K)} & $\wtwo$ (K) \\ \hline
\multirow{3}{*}{Base}   & $\mrm{R}^2$                    & \multicolumn{1}{c|}{0.05}   & \multicolumn{1}{c|}{0.76}  & 0.82  \\
                        & MAE                      & \multicolumn{1}{c|}{0.22}   & \multicolumn{1}{c|}{35}    & 35    \\
                        & RMSE                     & \multicolumn{1}{c|}{0.32}   & \multicolumn{1}{c|}{50}    & 56    \\ \hline
\multirow{3}{*}{CSO}    & $\mrm{R}^2$                    & \multicolumn{1}{c|}{0.19}   & \multicolumn{1}{c|}{0.56}  & 0.75  \\
                        & MAE                      & \multicolumn{1}{c|}{0.21}   & \multicolumn{1}{c|}{45}    & 43    \\
                        & RMSE                     & \multicolumn{1}{c|}{0.31}   & \multicolumn{1}{c|}{68}    & 63    \\ \hline
\multirow{3}{*}{CPD}    & $\mrm{R}^2$                    & \multicolumn{1}{c|}{0.35}   & \multicolumn{1}{c|}{0.79}  & 0.87  \\
                        & MAE                      & \multicolumn{1}{c|}{0.18}   & \multicolumn{1}{c|}{32}    & 30    \\
                        & RMSE                     & \multicolumn{1}{c|}{0.28}   & \multicolumn{1}{c|}{47}    & 45    \\ \hline
\multirow{3}{*}{FPD}    & $\mrm{R}^2$                    & \multicolumn{1}{c|}{0.37}   & \multicolumn{1}{c|}{0.82}  & 0.89  \\
                        & MAE                      & \multicolumn{1}{c|}{0.18}   & \multicolumn{1}{c|}{30}    & 28    \\
                        & RMSE                     & \multicolumn{1}{c|}{0.28}   & \multicolumn{1}{c|}{43}    & 42    \\ \hline
%Cerqueira \emph{et al.} & MAE                      & \multicolumn{1}{c|}{0.19}   & \multicolumn{1}{c|}{25}    & --    \\ \hline
\end{tabular}
\label{model_errors}
\end{table}

\section{Results}
\subsection{Database}

\begin{figure}[h!]
    \centering
    \includegraphics[width=\columnwidth]{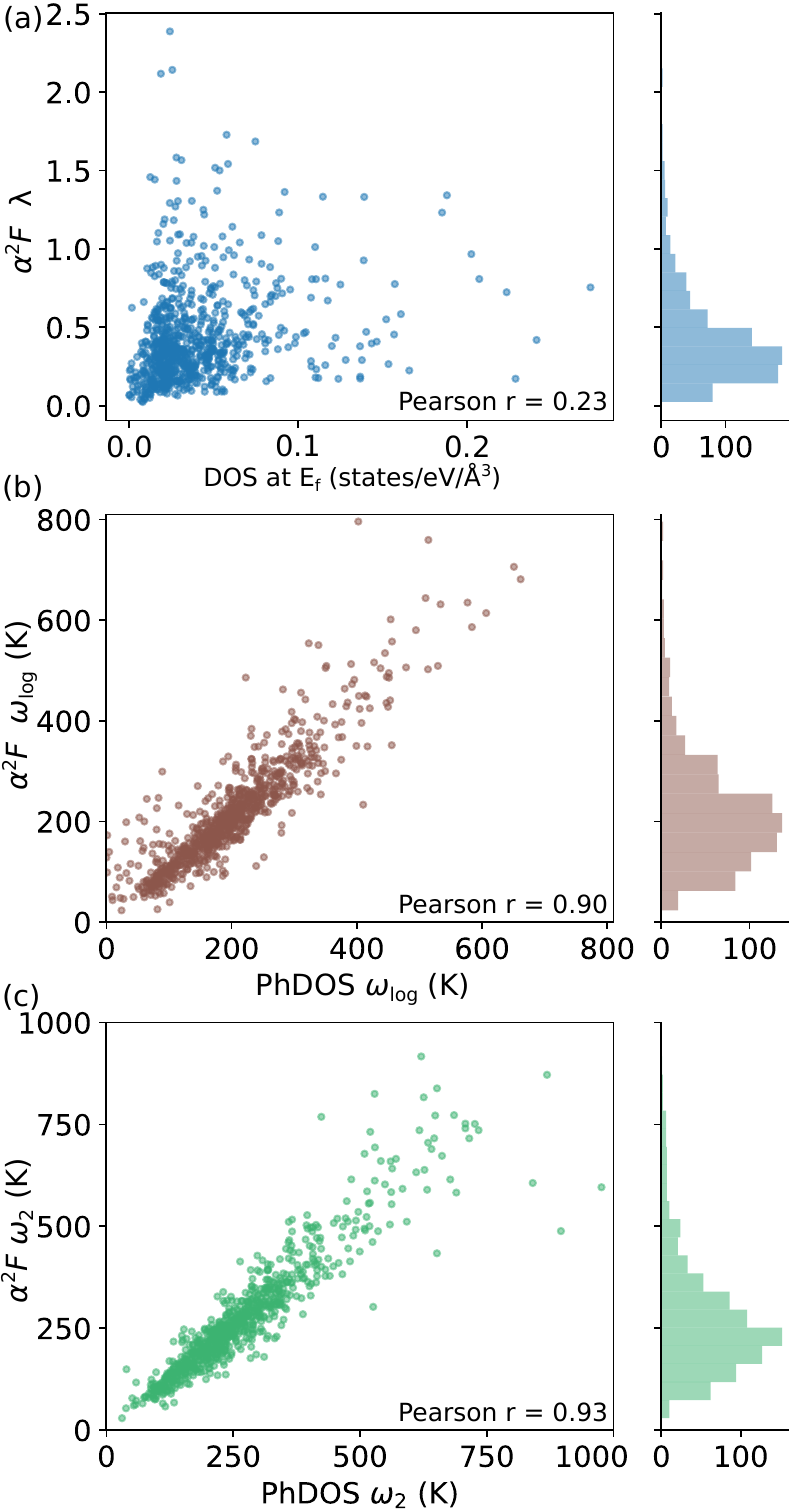}
    \caption{\textbf{Correlation and histogram plots of the dataset.} \textbf{(a)} Electron-phonon coupling constant, $\lambda$, and electronic density of states, eDOS, at the Fermi energy, \textbf{(b)} $\alpha^2 F$ and PhDOS derived $\omega_{\text{log}}$, and \textbf{(c)} $\alpha^2 F$ and PhDOS derived $\omega_2$. The histograms on the right-hand side show the distribution of the electron-phonon coupling constant and the two moments of $\frac{2}{\lambda\omega}\atf$ -- logarithmic moment $\omega_{\text{log}}$ and second moment $\omega_2$ -- in our dataset. The electronic DOS is averaged over a window of $\pm$50 meV at the Fermi energy. These correlation plots will guide the fit of a baseline model against which the deep learning models will be compared.}
    \label{base_model}
\end{figure}

We first analyze our DFT-calculated $\atf$ to develop a baseline model against which we will compare the performance of our deep learning models. For standardizing the choice of $\bk$ and $\bq$-grid used in the $\atf$ calculations, we develop an algorithm to generate grids based on user-provided $\bk$ and $\bq$-point densities, as described in the Methods section.
Figure~\ref{base_model} shows the correlations between the electron-phonon coupling constant, $\lambda$, and the electronic density of states, eDOS, at the Fermi energy and the correlation between the Allen-Dynes moments of $\alpha^2F(\omega)$~\cite{Allen-Dynes1975}, $\omega_{\text{log}}$ and $\omega_2$, and the coarse PhDOS for 818 materials in our database. Figure~\ref{base_model}(a) illustrates the very weak correlation between eDOS and $\lambda$, exemplifying that materials with a high eDOS do not always correspond to materials with a high $\lambda$. Approximately 72\% (595) entries have $\lambda$ less than 0.5, as high $\lambda$ often causes lattice instabilities~\cite{Pickett2023}.
In Fig.~\ref{base_model}(b) and (c), the moments of $\alpha^2F(\omega)$ have a high correlation with the corresponding moments of the coarse PhDOS. Allen and Dynes used $\omega_{\text{log}}$ derived from the PhDOS instead of $\omega_{\text{log}}$ from  $\atf$ in their seminal work in 1975~\cite{Allen-Dynes1975} based on the observed similarity between measured $\atf$ and PhDOS for tantalum. Here, we show that the similarity between the moments of $\atf$ and PhDOS extends beyond simple metals. This strong correlation will be used in our deep learning to enhance the model predictions.

To obtain the baseline model, we fit a multivariate linear regression model to the data presented in Fig.~\ref{base_model} that predicts the electron-phonon coupling constant $\lambda$, and the moments $\omega_{\text{log}}$ and $\wtwo$ from the eDOS and coarse PhDOS. Table~\ref{model_errors} summarizes the error metrics of the baseline model. The predictions from the baseline model are then used to calculate $\Tc$ using the Allen-Dynes equation~\cite{Allen-Dynes1975}, resulting in an MAE of 2.84~K. To put this MAE into perspective, Cerqueira \emph{et al.}~\cite{Cerqueira2023} and Choudhary~\cite{Choudhary2022} obtained an MAE of 2.94 and 1.39~K, respectively, using deep learning models that required crystal structure only as input. We further note that our baseline model for $\omega_{\log}$ has a lower testing MAE of 35~K compared to 37~K of the model by Choudhary~\cite{Choudhary2022} and is comparable to the cross-validation training error of 23~K for the model of Cerqueira \emph{et al.}~\cite{Cerqueira2023}.

From the coefficients of determination listed in Table~\ref{model_errors}, it is evident that our baseline model for $\lambda$ is marginally better than a `mean model' (a model that gives the mean of the training data for any testing data) and, in fact, the resulting $\Tc$ predictions are worse than the mean $\Tc$ of the training data ($\mathrm{R}^2 = -0.08$). Unfortunately, the previous two studies~\cite{Choudhary2022, Cerqueira2023} on predicting electron-phonon superconductivity do not provide $\mrm{R}^2$ and only evaluate the model performance using MAE. Using just one error metric to assess models can often overestimate their performance, as our baseline model shows for $\Tc$. Thus, we emphasize that at a minimum $\mrm{R}^2$, MAE, and RMSE must be reported for a fair assessment of regression models.

\subsection{Learning \texorpdfstring{$\alpha^2F(\omega)$}{a2F}}

\begin{figure*}
    \centering
    \includegraphics[width=0.70\textwidth]{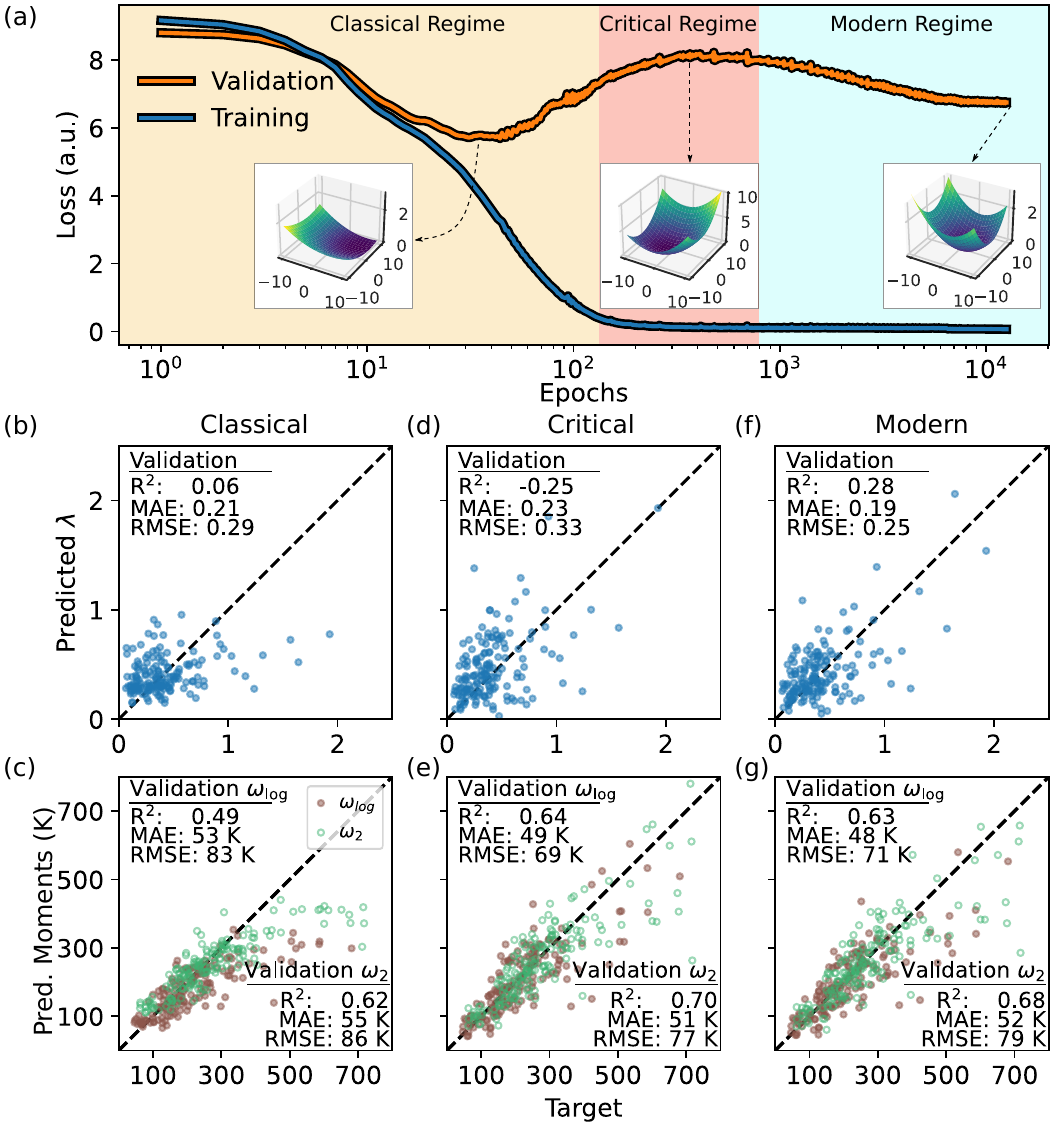}
    \caption{\textbf{Comparison of the classical, critical, and modern training regimes for a select model.} \textbf{(a)} The smoothed learning curve for the training (blue curve) and validation (orange curve) as a function of training epochs. The thickness of the line corresponds to small fluctuations in the loss. The insets show the loss landscape for the classical (1st minima), critical (maximum), and modern (2nd minima) regimes, illustrating why generalization improves for the second descent. The x and y-axes of the loss landscape are the magnitude of perturbation for each orthogonal direction, and the z-axis is the magnitude of loss. \textbf{(b, c)} The model trained in the classical regime's validation parity plots of $\lambda$, $\omega_{\log}$, and $\wtwo$, respectively. \textbf{(d, e)} The model trained in the critical regime's validation parity plots of $\lambda$, $\omega_{\log}$, and $\wtwo$, respectively. \textbf{(f, g)} The model trained in the modern regime's validation parity plots of $\lambda$, $\omega_{\log}$, and $\wtwo$, respectively.}
    \label{LC_props}
\end{figure*}

We partition the 818 dynamically stable materials into an 80-20 train-test split, ensuring a comparable representation of elements between the training and testing sets. The $\atf$ were binned and smoothed as outlined in the Methods section following a procedure similar to Chen \emph{et al.}~\cite{Chen2021_ENN}. BETE-NET is designed to predict the smoothed $\atf$ in bins of frequencies from 0.25 to 100.25 meV with a bin width of 2 meV. For learning the $\atf$, we choose the equivariant neural networks as they have repeatedly performed well on limited data due to their innate ability to identify unique motifs given a single observation~\cite{Frey2023, Rackers_2023, owen2023complexity}. Chen  {\it et al.}~\cite{Chen2021_ENN} provided a detailed GitHub repository of their equivariant neural network that achieved impressive PhDOS predictions for a dataset of comparable size. In our work, we adopted a modified version of their network (shown in Fig.~\ref{network}) and trained three variants. 

To evaluate the performance of the models, we compare the three properties $\lambda$, $\wtwo$ and $\omega_{\text{log}}$ derived from the predicted $\atf$ to the properties calculated from the DFT $\atf$.
During the training of the models, we noticed that the network initialization substantially impacts the models' performance, with the same model trained over different initializations producing highly variable validation results. To address this, we used bootstrapping~\cite{Bootstrap_Reference}, which inherently reduces sensitivity to initialization by averaging the effects of different initializations across multiple resampled datasets. Bootstrapping consists of generating new training sets of equal size to the original training set by sampling the original training set with replacement.  An ensemble is then generated by training multiple models on the bootstrapped datasets. An additional benefit of bootstrapping is that each resampled training set consists of only 62\% unique datapoints, retaining the remaining 38\% for validation.

A further dilemma of the limited data is the tendency for models to overfit quickly.  While classical machine learning considers overfitting detrimental to model generalization, many deep-learning models are trained to near-zero loss and maintain good generalization errors. This phenomenon is called double descent and can be viewed as controlled overfitting~\cite{overfitting}. Double descent is characterized by an initial decrease in validation loss to a minimum (classical regime) where the bias-variance trade-off holds. If training continues beyond this minimum, the validation loss will increase to a maximum (critical regime) where training loss is near zero, i.e., bias is zero, variance is at a maximum, and the model has learned one interpolant. Further training decreases validation loss (modern regime) as the model learns multiple interpolants, i.e., bias remains at zero and variance decreases~\cite{double_descent}. 

\begin{figure*}
    \centering
    \includegraphics[width=\linewidth]{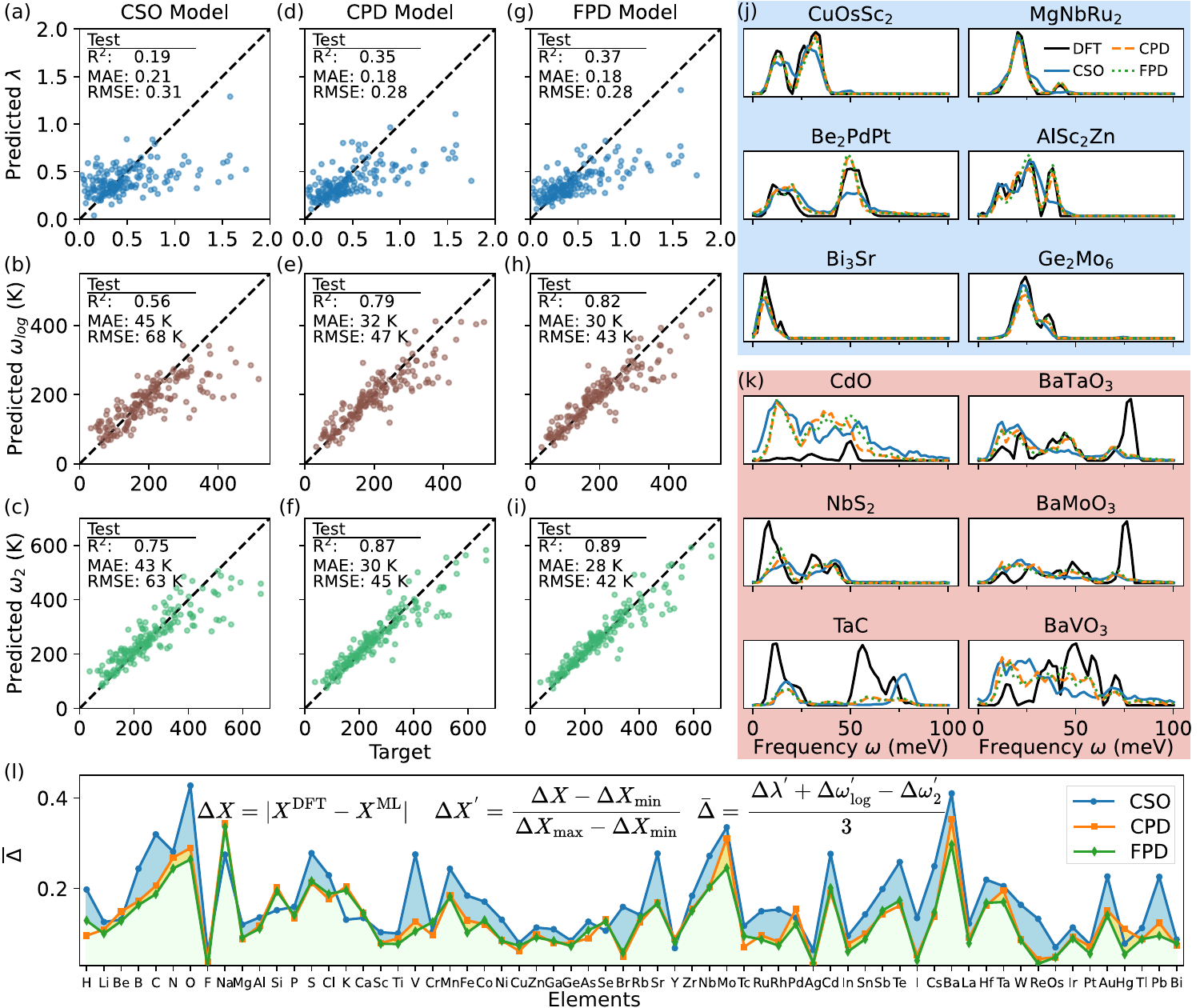}
    \caption{ \textbf{Test results for the three final variants of BETE-NET.} \textbf{(a-i)} Testing parity plots of $\lambda$, $\omega_{\log}$, and $\wtwo$, respectively, for the models. In comparing the parity plots across models, there is a systematic improvement in the derived properties, highlighting the advantage of embedding physically relevant properties. \textbf{ (j)} the six best predicted $\atf$. Here, we clearly see that embedding PhDOS almost perfectly corrects the predictions of the CSO model. \textbf{(k)} six worst predicted $\atf$. \textbf{(l)} The average prediction errors $\bar \Delta$ of materials containing each element.}
    \label{test_preds}
\end{figure*}

The three regimes are distinctly shown by the training and validation curves of Fig.~\ref{LC_props}(a).
We evaluated the validation prediction results of these regimes for a select model from our ensemble that best illustrates the improvement of the second descent. The model trained until the classical regime (Fig.~\ref{LC_props}(b, c)) obtains the lowest validation loss but only makes moderately accurate predictions for $\omega_{\text{log}}$ and $\wtwo$ and poor predictions for $\lambda$. We speculate that our choice of the loss function (mean squared error) allows a trivial solution of under-predicted peaks at sparsely represented frequencies. This under-prediction only yields a minimal penalty for the loss, leading to the classical model's poor performance for materials with high magnitude of the derived properties $\lambda$, $\omega_{\text{log}}$ and $\wtwo$. The model trained until the critical regime (Fig.~\ref{LC_props}(d, e))  performed much worse on $\lambda$ but obtained surprisingly better predictions on $\omega_{\text{log}}$ and $\wtwo$. Finally, the best overall performance was observed for the model trained until the modern regime (Fig.~\ref{LC_props}(f, g)). Like the model trained until the classical regime, these models are at a minimum validation loss but no longer learning the trivial solution. Instead, it learns to interpolate between training observations~\cite{overfitting}.

Providing a rationale for the improved performance of the second descent is an active area of research~\cite{Ba2020Generalization}. Similarly, machine learning theorists are actively debating how to correctly decompose the bias and variance of deep-learning models~\cite{yang2020rethinking}. To provide some understanding, we have visualized the loss landscape of the network using a simple qualitative approach proposed in Ref.~\onlinecite{losslandscape}. This method calculates the training loss after perturbing the learned network weights in two random orthonormal directions.
The loss landscapes for each of the three regimes are depicted in the insets of Fig.~\ref{LC_props}(a). The variance is represented by the curvature of the loss landscape as it describes the model's sensitivity to changes in the training set. We speculate that the distance between the minimum of the landscape and the initial model point (0,0) is the model's bias.
The model trained until the first minimum has the characteristic of the classic bias-variance trade-off. Visually, it has moderate curvature (variance), and the minimum of the loss landscape is moderately close to the initial model (moderate bias). The model trained until critically overfit has no bias (the landscape's minimum coincides with the unperturbed model) and high variance (curvature). Finally, the model trained until the second minimum maintains zero bias, but the curvature is much lower. This provides a visual explanation of the improved generalization of the second descent. 

We trained an ensemble of 100 bootstrapped models until the second descent to obtain BETE-NET. These models were validated using the data not sampled for a given bootstrapped training set, which corresponds to about 38\% of the total training set. We then applied the models to the test set, taking the final prediction as the average $\atf$ predicted by the ensemble. The CSO model (Fig.~\ref{test_preds}(a-c)) outperformed both the base model and the mean model. 

To further test BETE-NET on unseen data, we also apply the CSO variant to the 6,475 materials in the Cerqueira \emph{et al.}~\cite{Cerqueira2023} database. We used different DFT parameters to generate our dataset as compared to the one used by Cerqueira \emph{et al.}, resulting in slightly differing $\atf$, $\lambda$, and moments for the same materials (See Supplementary Fig.~1). One would expect a degradation in CSO's performance when applied to this bigger dataset. However, as seen in Supplementary Fig.~2(a), the CSO model gives error metrics similar to those reported in Fig.~\ref{test_preds}. Further, when comparing to the MAE from the model trained by Cerqueira \emph{et al.}, we find that the MAE in Supplementary Fig.~5(a) is, in fact, lower for $\lambda$ and similar for $\omega_{\log}$. This illustrates that our model architecture and training methodology are more data efficient and robust than existing models.

The CPD model (Fig.~\ref{test_preds}(d-f)) substantially improves the predictions for $\lambda$, $\omega_{\log}$ and $\wtwo$. This improvement is a product of embedding the coarse PhDOS, and as shown in Fig.~\ref{base_model}(b-c), the moments of $\atf$ are highly correlated to the moments of PhDOS. While this model requires some DFT calculations, the cost is substantially lower than a full $\atf$ calculation, yielding this model an excellent candidate as a secondary filter in a high throughput screening. This model substantially outperforms the base model and is the optimal model to directly compare to the base model as both contain information on the coarse PhDOS. Finally, the model containing information on the fine PhDOS (Fig.~\ref{test_preds}(g-i)) showed only marginal improvements over the coarse PhDOS model. We speculate that the coarse PhDOS captures the most relevant information for calculating $\atf$.

Next, we turn to the $\atf$ predictions to better understand BETE-NETs' performance and inner workings. Fig.~\ref{test_preds}(j) and (k) show the six best and worst $\atf$ predictions, respectively, decided based on the the prediction errors $\bar{\Delta }$ averaged over $\lambda$, $\omega_{\log}$, and $\wtwo$ for our test set. We calculate the $\bar{\Delta }$ by taking the mean of the normalized absolute difference between the DFT and ML predicted $\atf$ derived properties, as shown in the inset of Fig.~\ref{test_preds}(l). From the six best $\atf$ predictions for the test set, the CSO model predicts an $\atf$ that closely follows the DFT $\atf$. Adding information about the site-projected PhDOS (CPD and FPD models) further improves the prediction by guiding the models toward the expected shape of $\atf$. For example, for CuOsSc$_2$ and MgNbRu$_2$, the CSO model predicts extra peaks at 50 and 30 meV that the CPD and FPD models rectify. Furthermore, we suspect that by having access to the expected $\atf$ shape, the learning task is simplified for the CPD and FPD models, yielding better performance with regard to predicting the magnitude of the spectral function resulting in improved electron-phonon coupling constant prediction, as seen from the improvement of the error metrics for $\lambda$ (Fig.~\ref{test_preds}(a,d, and g)). Even so, the models tend to under-predict the magnitude of peaks at low frequencies, yielding under-predictions of $\lambda$ for materials with high magnitude $\lambda$.

We also attempt to understand the performance of BETE-NET on a per-element basis by plotting $\bar{\Delta }$ as a function of elements, as shown in Fig.~\ref{test_preds}(l). All our models perform moderately for materials that contain C, N, and O (Fig.~\ref{test_preds}(l)). The $\atf$ presented in Fig.~\ref{test_preds}(k) provides some physical insights on the performance of the models for materials containing these elements. The models struggle when predicting $\atf$ for phonon modes involving the light elements (C, N, and O). However, we prescribe caution in drawing strong conclusions from Fig.~\ref{test_preds}(l) with respect to expected trends in the per-element performance of our models due to the small test set. For example, 4 out of 5 Ba-containing materials also contain O. Likewise, all except one material with Mo contain either C, N, or O, leading to artificially high $\bar{\Delta}$. We direct the readers to Supplementary Fig.~2(b) for a better understanding of the per-element performance of our models, where we test the CSO model on 6475 materials from the Cerqueria \emph{et al.}~\cite{Cerqueira2023} dataset. Further improvements are expected by including materials containing light elements (H, B, C, N, and O), chalcogenides, and alkali metals in the training dataset. The small $\bar {\Delta}$ for transition elements (Supplementary Fig.~2(b)) implies that predictions of the CSO model are particularly robust for materials containing transition elements.

\subsection{Screening for High \texorpdfstring{$\Tc$}{Tc} Materials}

\begin{figure}
    \centering
    \includegraphics[width=\columnwidth]{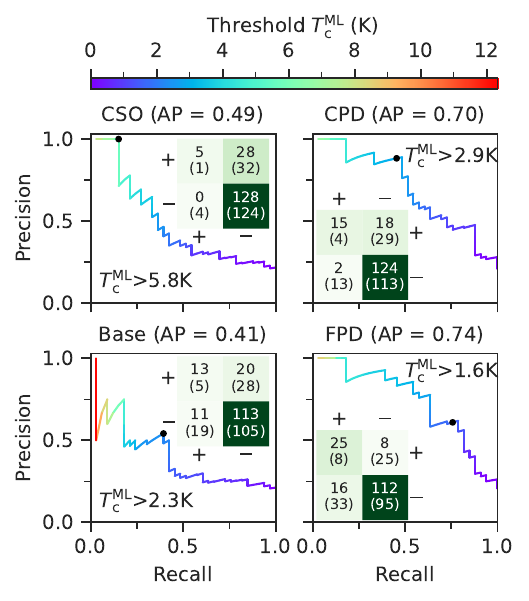}
    \caption{\textbf{Results of screening the test set for the 33 superconductors with $\Tc^{\mrm{DFT}}>5$~K.} The precision-recall curve for each model. Precision refers to the fraction of materials correctly predicted by our model to have $\Tc^{\mrm{DFT}}>5$~K compared to all predictions that meet this criterion. Recall refers to the portion of material correctly identified to have $\Tc^{\mrm{DFT}}>5$~K compared to all materials that meet this criterion. The color of the line represents the $\Tc^{\mrm{ML}}$ criterion. The title signifies the model, and the number in parentheses is the average precision (AP). The solid black marker signifies our suggested criterion for using the models as a surrogate model in high-throughput screening; the inset shows the confusion matrix for this criterion. The x-axis of the confusion matrix is the predicted label, the y-axis is the true label, and the numbers in parentheses show a comparison with a random classifier.  Compared to a random classifier, which would obtain an AP of 0.2 on our test set, all our models have better performance, with our best models (CPD and FPD) obtaining an AP nearly five times that of a random classifier. Our models perform better than the base model, which requires information on the PhDOS and eDOS.}
    \label{class}
\end{figure}

This section investigates BETE-NETs' ability to accelerate the search for high-$\Tc$ superconductors by first computing the Allen-Dynes critical temperature from the predicted and DFT computed $\atf$ for our test set. 
%\sout{and}
We then define all materials with $\Tc^{\mrm{DFT}} \geq 5$~K as high-$\Tc$ materials; 33 materials met this criterion. Rather than taking the naive approach of selecting $\Tc^{\mrm{ML}} \geq 5$~K, we compute precision and recall as a function of $\Tc^{\mrm{ML}}$. The precision and recall curve (Fig.~\ref{class}) provides a statistically robust analysis of the model's performance and a graphical way to balance the computational cost of high-throughput screening with the number of overlooked materials. To demonstrate this, we apply our models to the test set similarly to how they will be applied in high-throughput screening. 

First, we apply our CSO model to the data. In this initial screening, we can evaluate the materials at essentially no cost. However, given that the subsequent model will need DFT computed properties, we want to maximize the precision so only the most promising materials incur the cost of DFT computation. A criterion of $\Tc^{\mrm{ML}} \geq 5.8$~K gives a precision of 1.0 and a recall of 0.15. This distinct advantage of maintaining perfect precision is depicted by the black marker in the CSO model's precision-recall curve. While we overlook 85\% of the target material for this criterion, every material we evaluate with DFT has a $\Tc^{\mrm{DFT}} \geq 5$~K. Opting for perfect precision at the cost of low recall is beneficial when evaluating tens of thousands of materials, as we can quickly narrow our search to the most promising materials.

With the most promising materials identified by the CSO model, the CPD model can further reduce the number of candidate structures. Given our limited test set, we apply our CPD model to the full test set to illustrate its improved accuracy. At this point, we will have already incurred the expense of computing the coarse PhDOS and identified whether the material is dynamically and thermodynamically stable. As such, a lower precision is acceptable to obtain a larger recall. We select our criterion as $\Tc^{\mrm{ML}} > 2.9$~K illustrated by the black marker in the CPD model's precision-recall curve. At this criterion, we have a recall of 0.43 and a precision of 0.88, meaning we identified 43\% of the materials in our test set with $\Tc^{\mrm{DFT}} > 5$~K, and of those predicted materials, 88\% will be high-$\Tc$ materials. We provide the precision-recall curves for the base and FPD models for comparison.

Other studies~\cite{Cerqueira2023, Choudhary2022} only compared their model predictions directly to DFT, as theoretical predictions often diverge from experimental observations. Still, the ultimate test of the utility of BETE-NET is whether its predictions extend to experimental observations. As such, we apply our CSO model to experimentally stable metals in the Materials Project database~\cite{Jain2013}, removing metals that had the same composition as our dataset, yielding 11206 materials, of which 88 exhibited a $\Tc^{\mathrm{ML}} \geq 5.8$~K. Of the 88 materials, we identified 6 materials that had a documented $\Tc^{\mathrm{exp}}\geq 5$~K (see Supplementary Table I). Since identifying even a single superconductor is a monstrous task, the correct identification of 6 materials highlights the promise of our model in real-world application. We intend to investigate the remaining 82 materials predicted by our model in a proceeding study.

\section{Discussion}
%{\color{mag}Structure for this section based on the CHGNET paper}

Deep-learning methods have profoundly impacted many aspects of science and society. However, the limited data availability in some areas of the physical sciences has hindered the adoption of traditional deep-learning methods. This limitation necessitates the development of innovative ML techniques that can incorporate physical knowledge directly into the models, thereby efficiently utilizing limited data. Superconductivity research, where identifying novel superconductors can revolutionize electronics, power transmission, and magnet technology, is an exemplary field for adopting such a method. This is because identifying novel superconductors using first-principle methods alone is an extremely complex task, and developing sufficiently large datasets is prohibitively expensive at present. Addressing these challenges, our work presents BETE-NET, which integrates domain-specific knowledge and unconventional deep-learning techniques to temper overfitting, providing a robust approach to the difficult task of learning the Eliashberg spectral function using a small dataset.

In summary, we generated a database of 818 high fidelity $\atf$ calculations using our algorithm for standardizing the choice of $\bk$ and $\bq$-grids. We trained a Bootstrapped Ensemble of Tempered Equivariant graph neural NETworks (BETE-NET) using bootstrapping to improve generalization and reduce sensitivity to the network's weight initialization. We leveraged the double descent phenomenon - which, to the best of our knowledge, has not been discussed in the context of material science - to enhance predictions by allowing tempered overfitting. By plotting the loss landscape, we presented a plausible methodology to visually interpret neural networks' bias and variance, providing a qualitative rationale for the double descent phenomenon. Further, our predictions are enhanced by the inductive bias imposed by embedding the site-projected PhDOS. Finally, we showed how our models could act as a surrogate to screen for high-$\Tc$ materials, with our best model obtaining an average precision nearly five times that of a random selector.

Even with these methodical and architectural advances, the prediction of BETE-NET can be further improved. Specifically, we will supplement our database to include more examples of materials with relatively high electron-phonon coupling constants and materials containing light elements. Additionally, enforcing linear dependence on adjacent bins of the predicted $\atf$ can likely enhance the results. Linear dependency can be imposed by utilizing alternative loss functions such as a modified Wasserstein distance~\cite{EMD} or a curvature penalty similar to the penalty discussed by Xie~\emph{et al.}~\cite{UF3}. Alternatively, given a larger dataset, our strategy of fitting a bootstrapped ensemble to the second descent will allow a finer bin resolution yielding predicted $\atf$, which matches more closely with the raw DFT $\atf$, as the method utilizes data efficiently.

In conclusion, BETE-NET exemplifies the integration of domain-specific knowledge and advanced deep-learning techniques, offering a novel approach in superconductivity research to efficiently predict the Eliashberg spectral function using limited data, thereby broadening the scope of computational exploration and potentially leading to transformative societal impacts through the discovery of new superconductors.

\section{Methods}
\subsection{Dataset Generation}
All the DFT calculations are performed in Quantum Espresso~\cite{qe1,qe2,qe3} with the PBEsol functional~\cite{Perdew2008}. All structures are first relaxed using a $\bk$-point density of 40~$\AA^{-1}$. For calculating the electron-phonon matrix elements and the isotropic Eliashberg spectral function, we use the interpolation scheme by Wierzbowska, \emph{et al.}~\cite{Wierzbowska_epj}. The calculations of electron-phonon coupling require that the $\bk$-point mesh for calculating Kohn-Sham wave-functions and the $\bq$-point mesh for calculating phonons be commensurate. To achieve commensurate meshes we set the $\bk$-point and $\bq$-point density to 40~$\AA^{-1}$ and 15~$\AA^{-1}$, respectively and calculate the number of subdivisions for uniform $\bk$ and $\bq$-grids. Relaxed lattice parameters are used for the grid generation, and the grids are evaluated for commensuration along each direction independently. If the grids are not commensurate, the number of subdivisions in each direction in the $\bk$-grid first increases by one; commensuration is checked, then decreases by one if not commensurate. If the grids are still not commensurate, then the subdivisions in the $\bq$-grid are reduced by one, and the process is repeated until q-grid reaches $1\times1\times1$ at which point an error is raised. We found that the algorithm described above of generating $\bk$ and $\bq$-grid  provides a roughly uniform density across materials while maintaining sufficient accuracy for small volume unit-cells and reducing the cost of computations in large unit-cells, in contrast to using the fixed-size $\bk$ and $\bq$-grids  for all materials.

The $\atf$ is calculated on a set of 30 smearing values, $\sigma$, for the double $\delta$ integration~\cite{Wierzbowska_epj} with a step size of 0.001 Ry. In our experience, the $\alpha^2 F(\omega)$ calculated using $\sigma = 0.02$~Ry produced spectral functions that match closely with the ones published in the literature for elemental materials (see Supplementary Fig.~3). The 'Fine PhDOS' was calculated by Fourier interpolating the force constants obtained during the electron-phonon calculations onto a $20\times20\times20$ grid. For the 'coarse PhDOS', we recalculate force constants on $2\times2\times2$ grid and Fourier interpolated onto a $20\times20\times20$ grid. In all the DFT calculations, we set the kinetic energy cutoff for wave-functions to 75 Ry, the density cutoff to 350 Ry, and use Methfessel–Paxton smearing with a smearing width of $0.02$ Ry. The optimized norm-conserving Vanderbilt pseudopotentials~\cite{Hamann2013, Schlipf2015} available at SG15~\cite{ONCVPseudo} were used for all the DFT calculations except for the eDOS calculations. The eDOS calculations were carried out by using the tetrahedron $\bk$-point mesh and the ONCV pseudopotential available at PseudoDojo~\cite{vanSetten2018} as the SG15 pseudopotential lacked the atomic projectors.

In addition to recalculating most of the $\atf$ from Ref.~\onlinecite{Choudhary2022}, we further carried out electron-phonon calculations on small unit cell metals ($\leq 8$ atoms per unit cell) randomly selected from the Materials Project~\cite{Jain2013}. We also performed these calculations for metals not marked as `experimental', if they had a hull distance of less than 50~meV/atom, and contained less than five element types. We submitted 1,600 $\alpha^2F(\omega)$ calculations out of which 1,265 converged. Of these converged results, 399 materials had imaginary phonon modes at a frequency of more than 1~meV/atom, 38 had negative values in the Eliashberg function, and files generated by QE had ’NaN’ values for four materials. The eDOS calculations revealed that six materials marked as metals had a band gap. In the final dataset consisting of 818 $\atf$, there are 27 unary, 456 binary, 333 ternary, and two quaternary materials that we split into training and testing as outlined in the Methods section.
%To get the coarse PhDOS, force constants calculated on a $2\times2\times2$ $\bq$-grid were Fourier interpolated onto a $20\times20\times20$ grid. 

\subsection{Deep-Learning Models}
\textbf{Data preparation:} From the 1265 materials in which we computed the $\atf$ we selected the 818 materials that were dynamically stable. We then partitioned this dataset into training (80\%) and testing (20\%). This partitioning was done such that elements had comparable representation in both training and testing data (Supplementary Fig.~4). The bootstrapped datasets were determined by randomly sampling the training data with replacement to produce 100 bootstrapped training sets of equal length. Following this sampling produces bootstrapped datasets that each contain 62\% unique datapoints. The remaining 38\% is used for validation. The DFT calculated $\atf$ written to files had an energy resolution of 0.1 meV. Predicting $\atf$ at this level of resolution would require the ML models to have a substantial number of learned parameters. Given our limited data and to ensure consistent resolution and output dimensions, we follow the procedure used to smooth the PhDOS in Ref.~\onlinecite{Chen2021_ENN}. That is, we applied a Savitzky-Golay filter of window length 101 and polynomial order 3 and interpolated the smoothed $\atf$ onto 51 points over an energy range of $0.25 \leq \omega \leq 100.25$ meV. This same process was applied to the embedding of the site-projected PhDOS. The smoothing of $\atf$ leads to a minimal loss in information (see Supplementary Fig.~5).
%It is important to note that we limit our model to predictions in this discrete range.
%, limiting its application to dynamically stable materials that do not have peaks above 100.25 meV. 

\textbf{Model Training and Optimization:}
We trained our model using a mean squared error loss function and the adamW optimizer implemented in PyTorch~\cite{PyTorch} with a fixed learning rate of 0.005 and no weight decay. The model had a cut-off radius of 4~$\AA$, an embedded feature length of 64, irreducible multiplicity of 32, 2 point-wise convolution layers, 10 radial basis functions with the radial network consisting of a single layer, 100 neurons wide. Further details of the Euclidean neural network (e3NN) architecture can be found in Refs.~\onlinecite{e3nn_1, e3nn_2, e3nn_3}. We trained BETE-Net for a maximum of $10^5$ epochs, stopping training only if the validation loss was not reduced for $5\times10^3$ epochs. The final prediction is the mean of the predicted $\atf$ from the bootstrapped ensemble.

\section{Code Availability}
Code for implementing and training the models will be available at \url{https://github.com/henniggroup/BETE-NET} once the paper is published. 
\section{Data Availability}
The data used for training and testing the model will be available at \url{https://github.com/henniggroup/BETE-NET} once the paper is published.
\section{Acknowledgments}
The authors acknowledge the helpful insights and discussion with Gregory R. Stewart, James J. Hamlin, Laura Fanfarillo, and the entire Superconductivity Discovery Team at the University of Florida. This work was funded by the U.S. National Science Foundation, Division of Materials Research, under Contract No.\ NSF-DMR-2118718. A.C.H.\ and R.G.H.\ acknowledge additional support from the National Science Foundation under award PHY-1549132 (Center for Bright Beams). Part of this research was performed while J.B.G., A.C.H., and R.G.H. were visiting the Institute for Pure and Applied Mathematics (IPAM), which is supported by the National Science Foundation (Grant No. DMS-1925919). Computational resources were provided by the University of Florida Research Computing Center.

\section{Author Contributions}
J.B.G.\ and A.C.H.\ contributed equally to this work. J.B.G., A.C.H., P.J.H., and R.G.H. conceived the project. A.C.H. performed all DFT calculations and database generation. J.B.G. performed all model implementation and training. A.C.H. and J.B.G. contributed equally to all design decisions and analysis of the model and the data. P.M.D. provided context into the theory of superconductivity. B.G. provided helpful comments and critiques on the analysis and strategy adopted while developing the project. O.B. helped assess the viability of alternate architectures. B.G., P.J.H., and R.G.H. supervised the research. J.B.G., A.C.H., P.M.D., P.J.H., and R.G.H. wrote the manuscript. All authors contributed to revising and editing the manuscript.

\bibliography{references}

\end{document}